\documentclass[aps,prb,reprint,unsortedaddress,superscriptaddress]{revtex4-1}

\usepackage{amsmath,amssymb}
\usepackage[dvips]{graphicx}
\usepackage{bm}
\usepackage{color}
\begin{document}

\title{Weak topological insulator with protected gapless helical states}

\author{Ken-Ichiro Imura}
\affiliation{Department of Quantum Matter, AdSM, Hiroshima University, Higashi-Hiroshima 739-8530, Japan}
\author{Yositake Takane}
\affiliation{Department of Quantum Matter, AdSM, Hiroshima University, Higashi-Hiroshima 739-8530, Japan}
\author{Akihiro Tanaka}
\affiliation{National Institute for Materials Science, Tsukuba 305-0047, Japan}

\begin{abstract}
A workable model for describing dislocation lines introduced into a three-dimensional 
topological insulator is proposed.
We show how fragile surface Dirac cones of a weak topological insulator
evolve into protected gapless helical modes confined to the vicinity of
dislocation line.
It is demonstrated that surface Dirac cones of a topological insulator 
(either strong or weak) acquire 
a finite-size energy gap,
when the surface is deformed into a cylinder penetrating the otherwise surface-less
system.
We show that when a dislocation with a non-trivial Burgers vector is introduced, 
the finite-size energy gap play the role of stabilizing 
the one-dimensional gapless states.
\end{abstract}

\date{\today}

\maketitle

\section{Introduction}
The topological insulator has become one of the cutting-edge paradigms of 
the condensed-matter community
since the last couple of years.
\cite{Moore_review, Hasan_Kane, Qi}
Especially highlighted is the $Z_2$ topological insulator, 
\cite{Kane_Mele_Z2}
which has a band gap generated by spin-orbit coupling, 
and preserves time-reversal symmetry.
Though the idea of $Z_2$ topological insulator stems from 
the two-dimensional (2D) quantum spin Hall effect,
\cite{Kane_Mele_QSH}
its three-dimensional (3D) counterpart
has given a stronger impact on material science, leading, in particular, 
to the reclassifying of thermo-conducting layered crystals such as Bi$_2$Se$_3$ 
and Bi$_2$Te$_3$ as "strong" topological insulators.
\cite{Hasan_Kane}
In contrast to its 2D analogue,
the 3D $Z_2$ topological insulator has both weak and strong phases.
\cite{Fu_Kane_Mele, Fu_Kane}
A strong (weak) topological insulator bears an odd (even)  number of surface
Dirac cones when it is in contact with the vacuum, 
and is characterized by a $Z_2$-invariant $\nu_0=1$ ($\nu_0=0$).
Full characterization of a 3D $Z_2$ topological insulator requires,
however, a set of in total four $Z_2$ numbers: $(\nu_0; \nu_1 \nu_2 \nu_3)$.

In contrast to the topological number $\nu_0$ that characterizes
a strong topological insulator (STI) and is associated with a protected 
surface single Dirac cone,
other "weak indices" are generally believed to be 
nonrobust quantities.
On a perfect lattice, 
this assertion is indeed justified.
A recent study, however, 
on the response of a topological insulator
to the introduction of
lattice dislocations, \cite{Ran}
e.g., screw and edge dislocations, 
suggests that such dislocation lines
{\it play the role of a probe}
for characterizing WTI,
in which both strong ($\nu_0$) and weak ($\nu_1 \nu_2 \nu_3$) 
indices come into play.
The authors of Ref. \cite{Ran} have shown that both WTI and STI,
when twisted by dislocations,
accommodate a pair of protected one-dimensional (1D) helical modes.
This seems to contradict 
the common belief that a WTI is not topologically robust.
It is also 
counterintuitive 
that there 
always appear an even number, say, two pairs of Dirac cones on the 2D surface of a WTI,
whereas along a dislocation the number of protected 1D Dirac cones is 
at most one. 
The former is susceptible 
to disorder, especially 
to inter-valley scattering by short-range impurities,
whereas the latter is spin-protected from scattering by non-magnetic impurities.

\begin{figure}
\begin{center}
\includegraphics[width=8cm]{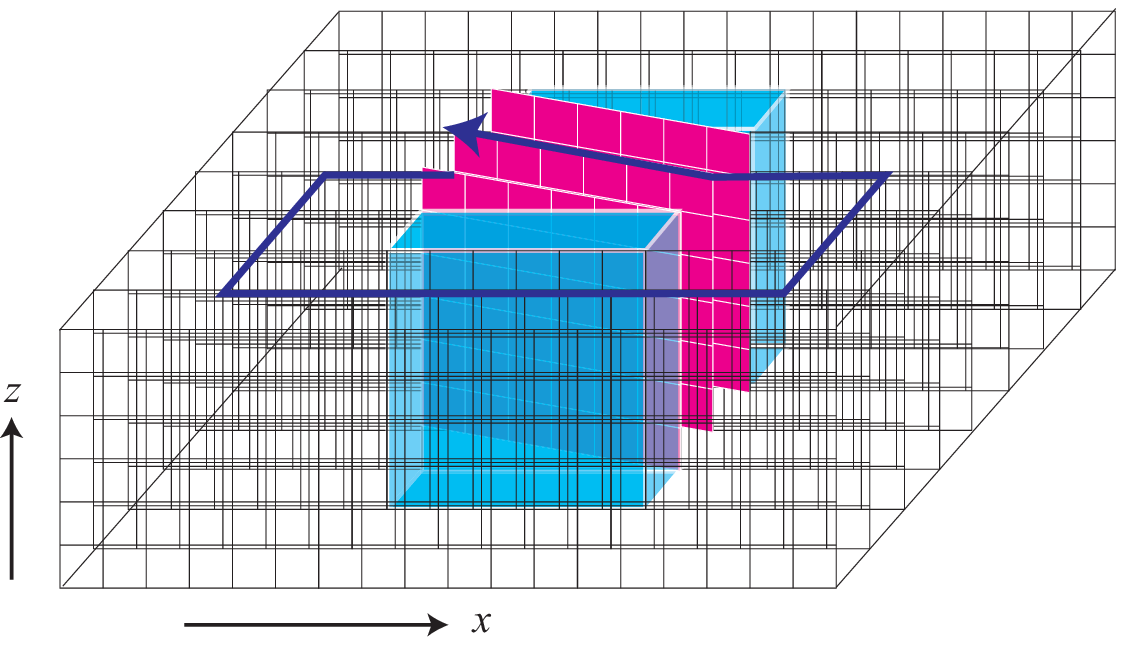}
\includegraphics[width=8cm]{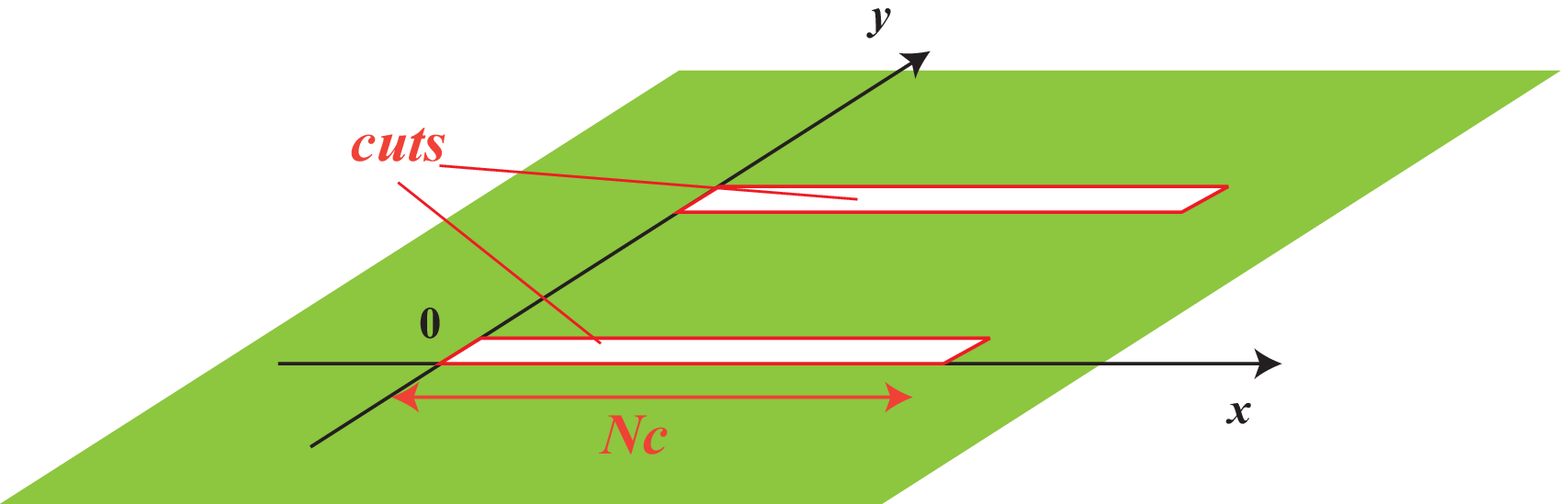}
\end{center}
\caption{A pair of screw dislocations (upper) 
with Burgers vector $\bm b = (0, 0, \pm b)$
inserted between the two cuts (lower).
The dislocation line is along the
$z$-axis and parallel to the Burgers vector.
The system is translationally invariant 
in the $z$-direction.
}
\label{screw}
\end{figure}

\begin{figure}
\begin{center}
\includegraphics[width=8cm]{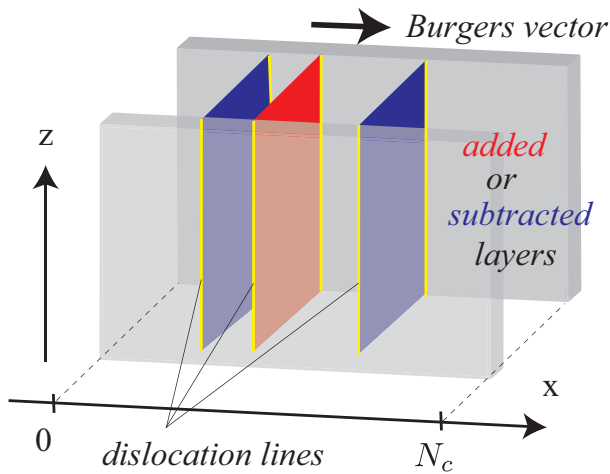}
\end{center}
\caption{Edge dislocations; a concrete implementation between the two cuts.
Here, the Burgers vector $\vec{b}$ is parallel to the $x$-axis;
$\vec{b}=(b, 0, 0)$.}
\label{edge}
\end{figure}

The aim of this paper is to resolve the above 
seemingly opposing points of view on the
behavior of WTI on a 2D surface and along a 1D dislocation line.
We propose a concrete theoretical model that is intended to interpolate 
between the two cases.
To implement either screw or edge dislocations;
see Figs. \ref{screw} and \ref{edge}), 
we first introduce two cuts extended in parallel with the $z$-axis.
For analytic considerations it is more convenient to regard
such linear cuts (of width $N_c$) as cylindrical punctures
(of circumference $s$)
penetrating the otherwise surfaceless system.
By "cuts" we mean links
on which electron hopping is switched off in the tight-binding
description.
A pair of screw (edge) dislocations are then introduced around (between)
these two cuts.
Electrons in the surface states 
(only such electrons are relevant to transport characteristics)
can be seen as a collection of 1D modes that 
come in pairs (Kramers' pair),
moving up and down the punctures.
These electrons also feel the existence of crystal
dislocations.
The latter plays a role similar to that of an (imaginary) magnetic
flux piercing the puncture.
The previously mentioned 2D and 1D cases are naturally included 
within this model as the limit of, respectively, $s\rightarrow\infty$ 
and $s\rightarrow 0$.
We follow the evolution of electronic states along such punctures
with a non-trivial lattice distortion as $s$ is varied.
It is revealed that the topological stability of protected 1D gapless
helical modes stems from a finite-size energy gap associated with
the spin Berry phase.
The latter has been a subject of much theoretical attention
\cite{Franz_wormhole, Ashvin, Moore}
in the context of peculiar Aharonov-Bohm oscillations
observed recently in a system of STI.
\cite{AB_exp}
The protected gapless modes along dislocation lines
have been studied also from the viewpoint of engineering
thermo-electric materials.
\cite{Murakami}

\section{Model}

In the bulk (outside the punctures and away from the dislocation)
we consider a lattice version of the following simplified model for
3D $Z_2$ topological insulator:
\cite{Franz_Witten,Franz_wormhole}
\begin{equation}
H = A k_\mu \gamma_\mu + 
(\Delta - B  k_\mu k_\mu) \gamma_0
\label{Dirac}
\end{equation}
where repeated indices should be summed over
$\mu=1,2,3$.
$\gamma$-matrices are chosen, e.g., as,
\begin{equation}
\gamma_\mu = \tau_z \sigma_\mu,\ \ \
\gamma_0 = \tau_x
\label{gamma}
\end{equation}
for $\mu=1,2,3$.
Then, following the same type of procedure as described in 
Refs. \cite{KI_PRB2010, Franz_Witten},
we place the model on 
a 3D square lattice of size $N_x \times N_y \times N_z$,
and impose, 
unless stated otherwise
a periodic boundary condition in each direction.

Away from the two cuts and dislocations,
our tight-binding Hamiltonian reads,
\begin{eqnarray}
H &=& \sum_{x,y,k_z} \big\{
m (k_z) |x,y,k_z\rangle \langle x,y,k_z| 
\nonumber \\
&+& \big(
t_x |x+1,y,k_z\rangle \langle x,y,k_z| 
\nonumber \\
&+& t_y |x,y+1,k_z\rangle \langle x,y,k_z| + h.c.
\big)
\big\}
\label{ham}
\end{eqnarray}
where
\begin{eqnarray}
m (k_z) &=& A \sin k_z \tau_z \sigma_z +
(\Delta - 6B + 2B \cos k_z) \tau_x,
\nonumber \\
t_x &=& i {A \over 2} \tau_z \sigma_x + B \tau_x,
\nonumber \\
t_y &=& i {A \over 2} \tau_z \sigma_y + B \tau_x.
\end{eqnarray}

\subsection{Cuts}

In order to implement a punctured geometry and introduce dislocations
on the square lattice, 
we first deform the punctures into the form of a "cut" 
(see the lower panel of FIG. \ref{screw})
of length $N_c$ (its circumference is $s=2N_c$).
We introduce two cuts, then 
a pair of screw (Fig. \ref{screw}) or edge (Fig. \ref{edge})
dislocations between them.
As shown in these figures, 
here the two cuts are placed along the $z$-axis,
and between the two crystal layers: $y=0$ and $y=1$
(as well as between $y=N_y /2 -1$ and $y=N_y /2$).
Between these crystal layers
hopping is turned off for $x=1,\cdots, N_c$.
Introduction of these two cuts breaks the discrete translational invariance
(crystal periodicity) in the $(x,y)$-plane,
whereas it preserves the translational invariance in the
$z$-direction, i.e.,
$k_z$ is still a good quantum number.
In the following, we will extensively investigate energy spectra: $E=E(k_z)$ 
of the system in the presence of screw or edge dislocations.

\subsection{Screw vs. edge dislocations}
\subsubsection{Case of screw dislocations}

A pair of screw dislocations can be introduced between the two cuts
by dislocating the hopping matrix elements in the region between the two cuts
(Figs. \ref{screw}),
i.e., for $y=1, \cdots, N_y/2$, say,
between the two crystal layers
$x=0$ and $x=1$ as
\begin{equation}
t_x |x+1,y,z\rangle \langle x,y,z| \rightarrow
t_x |x+1,y,z-b\rangle \langle x,y,z|.
\label{twist}
\end{equation}
$b$ measures the 
strength of the dislocation, i.e., the magnitude of the Burgers vector.
This is equivalent to "twisting" the same hopping matrix elements 
by a factor $e^{i k_z b}$
in the $k_z$-diagonalized basis.
\cite{Kawamura}
Note that the cuts and twist structure is translationally invariant in the $z$-direction,
and $k_z$ is still a good quantum number.

\subsubsection{Case of edge dislocations}

In the case of (a pair of) edge dislocations with burgers vector $\bm b = (\pm b, 0, 0)$,
we suppress $b$ nearest-neighbor hopping amplitudes in the $x$-direction 
between $x=x_0$ and $x=x_0+1+b$ and also between the two cuts ($1 \le y \le Ny/2$),
and instead introduce a "skipping" process,
\begin{equation}
t_x |x_0+1+b,y,z\rangle \langle x_0,y,z|,
\label{skip}
\end{equation}
to the tight-binding Hamiltonian (\ref{ham}).

\begin{table}
\caption{Three distinct topological phases of the
square lattice Dirac mode:
its low-energy effective Hamiltonian
around the $\Gamma$-point is given in
Eqs. (\ref{Dirac}), (\ref{gamma}).
}
\begin{center}
\begin{tabular}{lcccccccc}
\hline\hline
&\ \ \ & $\delta_0$ & $\delta_1$ & $\delta_2$ & $\delta_3$ &\ \ \ &
$(\nu_0; \nu_1 \nu_2 \nu_3)$& phase
\\ \hline
$\Delta /B<0$ && 
$+$  & $+$ & $+$ & $+$ && 
$(0; 000)$ & trivial
\\
$0<\Delta /B<4$ && 
$-$  & $+$ & $+$ & $+$ && 
$(1; 000)$ & STI
\\
$4<\Delta /B<8$ && 
$-$  & $-$ & $+$ & $+$ && 
$(0; 111)$ & WTI
\\
$8<\Delta /B < 12$ && 
$-$  & $-$ & $-$ & $+$ && 
$(1; 111)$ & STI
\\
$12 <\Delta /B$ && 
$-$  & $-$ & $-$ & $-$ && 
$(0; 000)$ & trivial
\\ \hline\hline
\end{tabular}
\end{center}
\label{indices}
\end{table}

\subsection{Strong vs. weak indices}

The 3D topological insulator model we employ
has three distinct topological phases as shown in Table 1.
Indices $\delta_0$, $\delta_1$, $\delta_2$, $\delta_3$
in the Table
are parity eigenvalues: $\pm 1$, respectively, at
$\Gamma = (0,0,0)$ (for $\delta_0$), at three 
inequivalent but symmetric $X$-points:
$(\pi, 0, 0)$, $(0, \pi, 0)$, $(0, 0, \pi)$ (for $\delta_1$),
at three $M$-points:
$(0, \pi, \pi)$, $(\pi, 0, \pi)$, $(\pi, \pi, 0)$
(for $\delta_2$)
and at $R=(\pi, \pi, \pi)$
(for $\delta_3$).
The above eight (by distinguishing inequivalent points) symmetry
points ($\Gamma$, $X$, $M$ and $R$) are also called
time-reversal invariant momenta (TRIM) of the 3D
Brillouin zone.
These parity eigenvalues are in our model related to the 
strong and weak indices as,
\begin{eqnarray}
(-1)^{\nu_0} &=& \delta_0 \delta_1 \delta_{1}' \delta_{1}'' \delta_2 \delta_{2}' \delta_{2}'' \delta_3
=\delta_0 \delta_1^3 \delta_2^3 \delta_3,
\label{nu_0} \\
(-1)^{\nu_1} &=&  \delta_1 \delta_{2}' \delta_{2}'' \delta_3 = \delta_1 \delta_2^2 \delta_3,
\label{nu_1} \\
(-1)^{\nu_2} &=& \delta_{1}' \delta_2 \delta_{2}''  \delta_3 =  \delta_1 \delta_2^2 \delta_3,
\label{nu_2} \\
(-1)^{\nu_3} &=& \delta_{1}'' \delta_2 \delta_{2}' \delta_3 =  \delta_1 \delta_2^2 \delta_3.
\label{nu_3}
\end{eqnarray}
Here, we have distinguished, for later convenience, three $\delta_2$'s and $\delta_3$'s 
at symmetric but inequivalent TRIM
(the value of these $\delta$'s are identical in our model with high symmetry;
as for definitions of these $\delta$'s, see Fig. \ref{3D_2D}).

In the following, we focus on the WTI phase:
$4<\Delta /B<8$,
and study how a protected 1D helical pair 
arises from a topologically fragile surface of a WTI.

\begin{figure}
\begin{center}
\includegraphics[width=8cm]{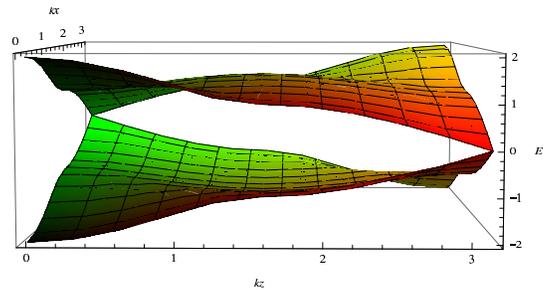}
\end{center}
\caption{Calculated spectrum of surface Dirac cones.
in the WTI phase ($\Delta/B = 6$, $A=B=1$). 
Periodic boundary condition in the $y$-direction is relaxed; 
the system 
forms a slab, or a torus of finite thickness.
The two Dirac cones are located at TRIM:
$(\pi, 0)$ and $(0, \pi)$.
}
\label{surf}
\end{figure}

\section{Energy spectrum of WTI in the presence of punctures and 
dislocation lines}

A WTI has an even number of Dirac cones on its surface
as depicted in Fig. \ref{surf}.
Here, the surface is chosen normal to the $y$-axis, i.e., 
a WTI occupying the half space $y<0$ is in contact with the vacuum
occupying the remaining half at the $y=0$ surface. 
The two Dirac cones are located at two
TRIM's:
$(0, \pi)$ and $(\pi, 0)$ 
in the surface coordinates $(k_z, k_x)$.

\begin{figure}
\begin{center}
\includegraphics{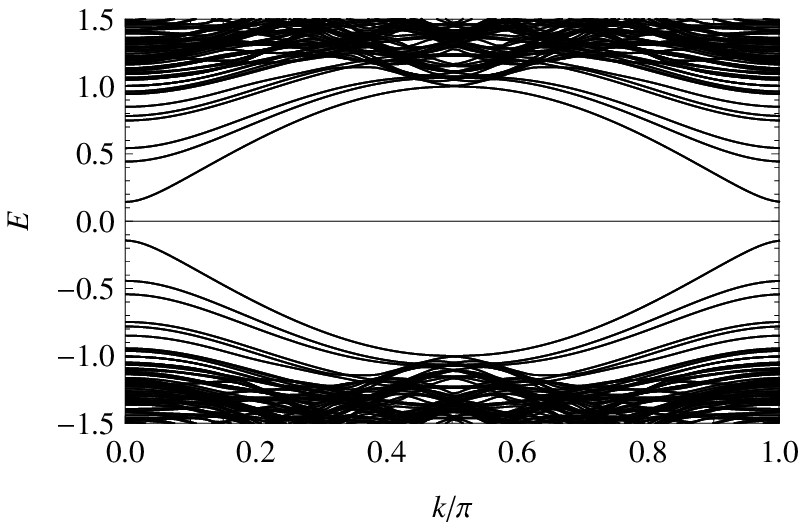}
\includegraphics{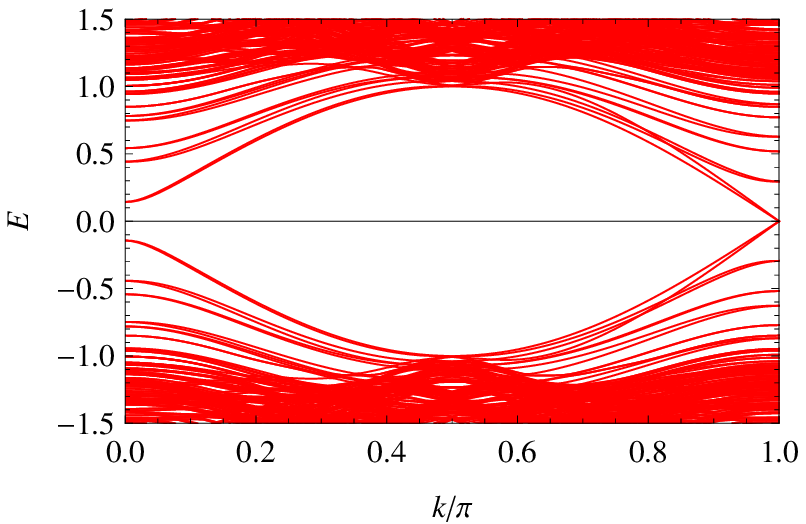}
\includegraphics{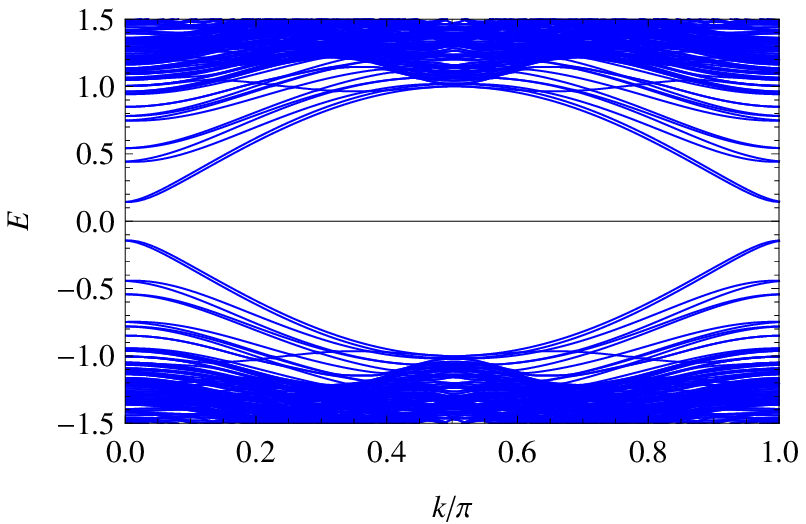}
\includegraphics{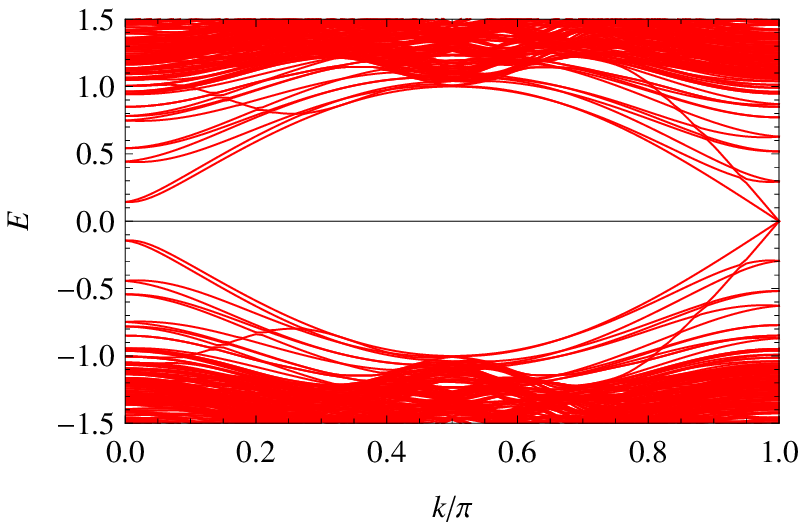}
\end{center}
\caption{Energy spectrum $E(k)$ of WTI ($\Delta/B = 6$)
in the presence of {\it screw} dislocations.
Here, the 1D momentum $k$ is chosen 
to be along the cuts;
$k=k_z$.
The Burgers vectors are gives as $\vec{b}=(0, 0, \pm b)$ with
$b=0$ (dislocation is absent), $b=1$, $b=2$ and $b=3$ 
respectively, in the top, second, third and bottom panels.
The calculation
is done for a system of size,
$N_x \times N_y = 16 \times 16$, and 
cut width $N_c = 8$. 
The other parameters are set as $A=B=1$.
}
\label{spec_screw}
\end{figure}

\subsection{Finite-size energy gap of surface Dirac cones on
a cylindrical surface}
Imagine 
deforming this flat surface into a cylindrical tube.
The tube is further deformed adiabatically into a cut of
Fig. \ref{screw} and Fig. \ref{edge}.
The two Dirac cones are now projected onto the $k_z$-axis,
as shown in the Fig. \ref{spec_screw}.
Notice that the two projected Dirac cones at
$k_z =0$ and $k_z = \pi$
have acquired a finite size gap in the 
upper 
panel.
Note that here the twist is {\it not introduced yet}.
The appearance of a gap is 
a rather unexpected phenomenon, if one recalls
that carbon nanotubes become either metallic or semiconducting
depending on the way a graphene is rolled up into a tube.
\cite{R}
Here, a crucial difference from the carbon nanotube case is
that the Dirac cone involves {\it a real spin} and not a sub-lattice pseudo-spin.
The procedure of rolling up a flat surface into a tube
introduces a $2\pi$-rotation in spin space along a contour
winding 
around the tube once.
\cite{Franz_wormhole, Ashvin, Moore}
The resulting $-1$ factor changes the boundary condition around 
the tube from periodic to anti-periodic:
\begin{equation}
e^{i (p_x + k^{(0)}_x) s} \times (-1) = 1.
\label{bc_0}
\end{equation}
Here, we have decomposed the total crystal momentum
of an electron 
into short- and long-wavelength components:
\begin{equation}
\bm k = \bm{k^{(0)} + p}.
\end{equation}
$\bm p = (p_x, p_y)$ 
refers to the long-wavelength component 
measured from the Dirac point.
The short-wavelength component 
$\bm k^{(0)} = (k^{(0)}_x, k^{(0)}_y)$ 
is, on the other hand,
a crystal momentum {\it at the Dirac point},
and typically $k^{(0)}_x = 0, \pi$.
Recall here that
the circumference $s = 2 N_c$ of the cut is, 
by its construction, an {\it even} integer multiple of the lattice constant,
since the cut is made
by disconnecting $N_c$ links of an otherwise 
locally perfect crystal.
This signifies that
\begin{equation}
e^{i k^{(0)}_x s} = 1
\label{bc_1}
\end{equation}
always holds.
As a result, the anti-periodicity of the boundary condition
(\ref{bc_0}) must be taken care of 
solely by the long-wavelength part of the crystal momentum, 
and eliminates, as seen e.g. in the spectrum of Fig. \ref{spec_screw} 
(top panel),  
states on the line $p_x = 0$ crossing the very
bottom of a Dirac cone.
Low-energy states in the same figure
consist of $p_x = \pm \pi /s$,
leading to occurrence of a finite-size gap,
\begin{equation}
\Delta E = 2A{\pi\over s}.
\label{gap}
\end{equation}
in the spectrum.

\begin{figure}
\begin{center}
\includegraphics{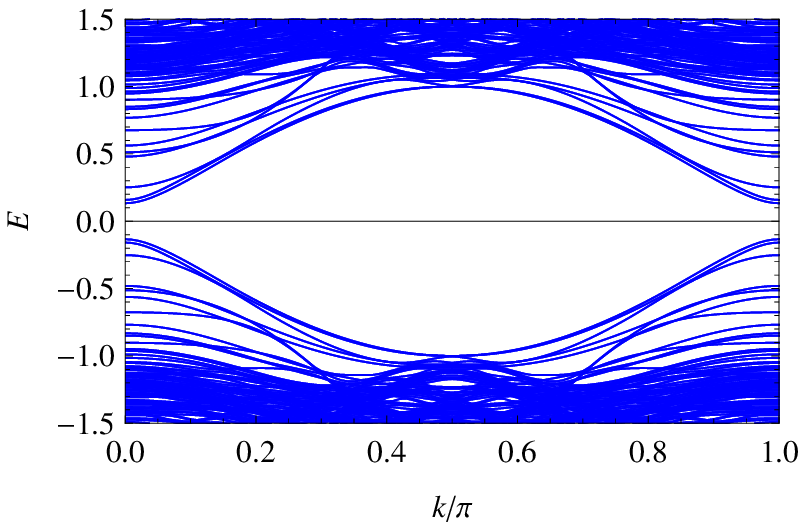}
\includegraphics{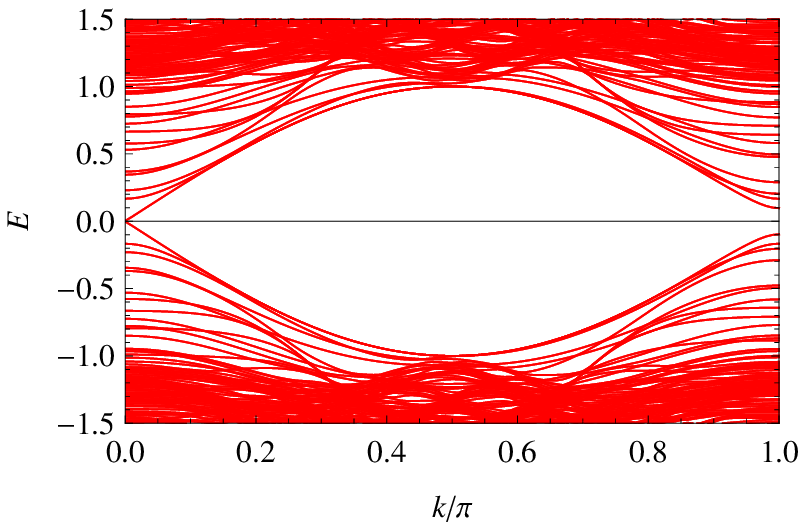}
\end{center}
\caption{Similar plots as the last two panels of Fig. \ref{spec_screw}
here in the case of {\it edge} dislocations.
The Burgers vector is here chosen as $\vec{b}=(\pm b, 0, 0)$ with
$b=2$ and $b=3$, respectively, in the upper and
lower panels.
}
\label{spec_edge}
\end{figure}

\subsection{Screw dislocations}
The second panel of Fig. \ref{spec_screw} shows, on the other hand,
the spectrum when the system is twisted by a pair of
screw dislocations
with Burgers vector $\vec{b} = (0, 0, \pm b)$
where $b=1$.
Such a lattice scale deformation modifies the periodicity of
the wave function associated with the
short-wavelength component of the crystal momentum,
i.e., $k^{(0)}_z= \pi$ in the present case.
Note that the entire effect of a {\it screw} dislocation can be concentrated on
hopping amplitudes across a single surface, as in Eq. (\ref{twist}).
Its influences on the electronic wave function sums up to
a phase shift 
$e^{i k_z b}$ 
on crossing the same surface
(here, this is a surface 
inserted between the two crystal layers $y=0$ and $y=1$).
Thus, adding this phase shift to (\ref{bc_0}) and
taking Eq. (\ref{bc_1}) into account,
one finds 
that the appropriate boundary condition in the presence
of a screw dislocation 
reads, 
\begin{equation}
e^{i p_x s} \times e^{i k^{(0)}_z b} \times (-1) = 1.
\label{bc_2}
\end{equation}
Note that here a small additional phase factor $e^{i p_z b}$
which modifies only 
{\it gapped} solutions with $p_z \neq 0$
has been omitted for the sake of clarity. 
Eq. (\ref{bc_2}) dictates that
only the surface Dirac cone projected onto 
$k^{(0)}_z = \pi$ is susceptible 
to the 
change of the magnitude of the Burgers vector, and closes the gap 
(i.e. the $p_x = 0$ state is now
allowed \cite{Ran_PRB})
when $b$ is an {\it odd} integer.

Some examples confirming this even/odd feature
are shown in Fig. \ref{spec_screw}.
In the last two panels of Fig. \ref{spec_screw},
one can also observe that 
Kramers pairs at $k_z =0$ exchange their partners as $k_z$ evolves
up to $k_z =\pi$,
in accordance with the twisting of boundary condition.

The anti-periodic boundary condition (\ref{bc_0}), the resulting finite-size gap (\ref{gap}),
as well as the twisting of the boundary condition such as Eq. (\ref{bc_2})
also underlie the origin of the anomalous Aharonov-Bohm oscillations 
observed recently in Bi$_2$Se$_3$ nanoribbbons.
\cite{AB_exp}
In the Aharonov-Bohm geometry, the twisting of the boundary condition 
\`a la Eq. (\ref{bc_2}) is caused, not by a dislocation,
but instead by a magnetic flux tube penetrating the puncture.
\cite{Franz_wormhole, Ashvin, Moore}

\subsection{Edge dislocations}

The above argument needs to be modified
in the case of an edge dislocation associated with the same dislocation line.
Such defects can be introduced 
e.g., as in Fig. \ref{edge},
in which dislocations terminate at a cut of finite
width similarly to the case of a screw dislocation.
The Burgers vector in this implementation is 
along the $x$-axis, $\bm b = (b, 0, 0)$.
Here, $b$ is the number of subtracted (added)
layers between the two cuts.
Recall that for an edge (screw) dislocation
the Burgers vector is perpendicular (parallel)
to the dislocation line (parallel to the $z$-axis here).
The effect of such an edge dislocation
on the electronic wave function
can be fully taken into account
as a change of the boundary condition
(\ref{bc_0}),
i.e., by the replacement: $s \rightarrow s+b$
in the same equation.
This leads, when (\ref{bc_1}) is taken into account, to 
\begin{equation}
e^{i p_x s} \times e^{i k^{(0)}_x b} \times (-1) = 1,
\label{bc_3}
\end{equation}
i.e., a twisted boundary condition analogous to Eq. (\ref{bc_2})
but with $k^{(0)}_z$ replaced by $k^{(0)}_x$.
Note 
that again there is a small additional phase factor $e^{i p_x b}$
which appears only when $p_x \neq 0$.
Eq. (\ref{bc_3}) dictates, in contrast to Eq. (\ref{bc_2}),
that among the two surface Dirac cones projected onto $k_z$-axis,
only the one with $p_x = \pi$ is susceptible 
to the presence of edge dislocation, and closes its finite-size gap 
when $b$ is an {\it odd} integer.
Two panels of Fig. \ref{spec_edge} indeed confirm
this even/odd feature in a few non-trivial cases: $b=2, 3$.
Notice that
protected gapless modes appear at $k_z=0$
[$b=3$ (odd) case],
in contrast to the case of screw dislocation.
This is because here
the underlying surface Dirac cone responsible for
the gap closing is located at $(k_z, k_x) = (0, \pi)$,
projected naturally to $k_z =0$.

\begin{figure}
\begin{center}
\includegraphics{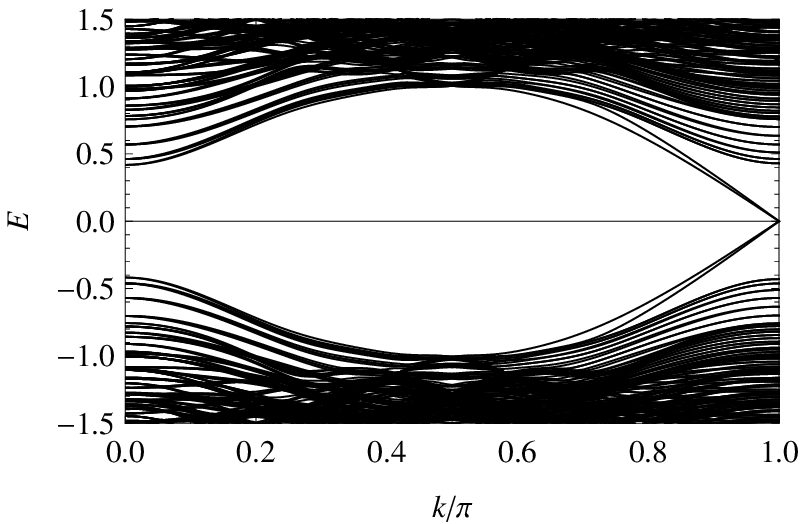}
\includegraphics{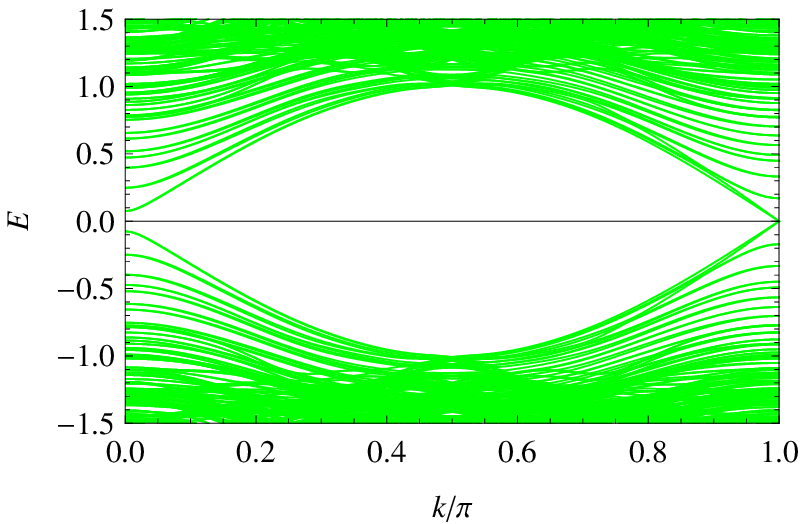}
\end{center}
\caption{Protected gapless helical modes along a pair of
{\it screw} dislocations ($b=1$).
The spectrum is calculated at $\Delta/B = 6$
for two different values of the cut-width $N_c$;
$N_c = 0$ (upper) and
$N_c = 16$ (lower). 
The size of the system is chosen as
$N_x \times N_y = 32\times 8$.
See the text for discussions.
}
\label{screw_c}
\end{figure}

\begin{figure}
\begin{center}
\includegraphics{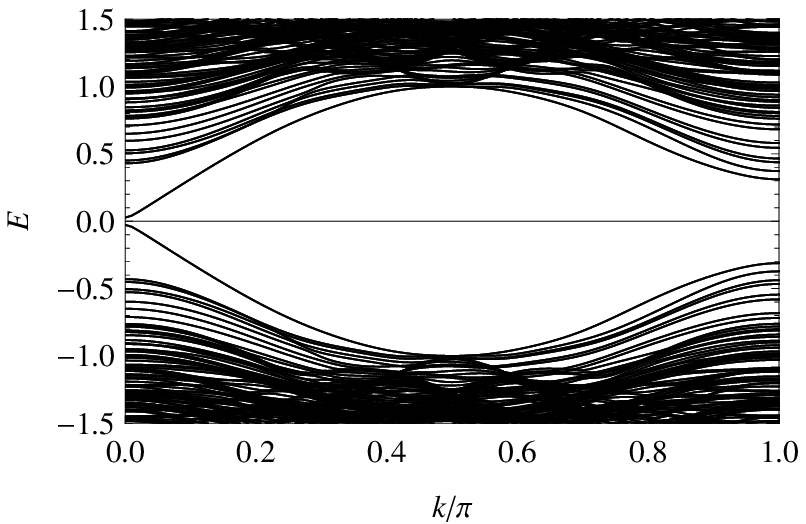}
\includegraphics{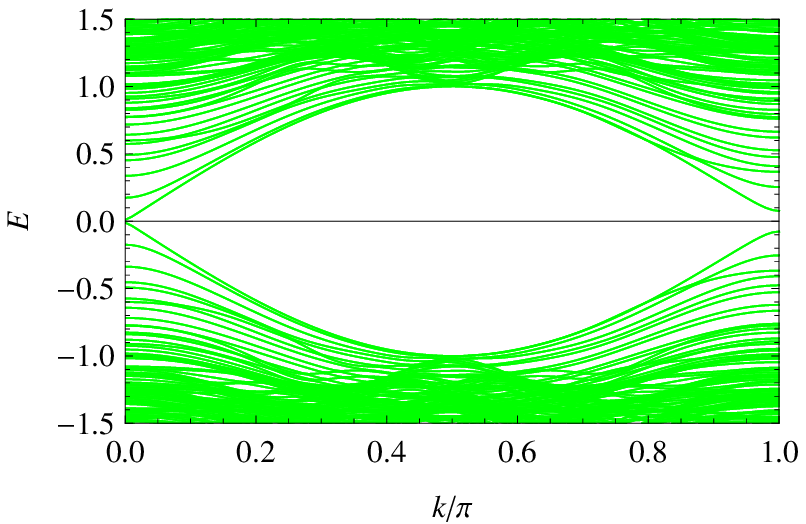}
\end{center}
\caption{Protected gapless helical modes along a pair of
{\it edge} dislocations ($b=1$).
The spectrum is calculated at $\Delta/B = 6$
for two different values of the cut-width $N_c$;
$N_c = 1$ (upper) and
$N_c = 16$ (lower). 
$N_x \times N_y = 32\times 8$.
}
\label{edge_c}
\end{figure}

\section{Finite size gap of projected 2D Dirac cones and the
protected 1D gapless helical modes}

We have seen that 2D surface Dirac cones 
attains a finite-size mass gap, when the surface is
deformed into a tube of finite circumference
$s$ (cf. Eqs. (\ref{bc_0}), (\ref{gap})).
We point out that this observation is the key for understanding 
the mechanism 
of how 
the originally fragile 2D surface Dirac cones of WTI
acquires {\it robustness} 
upon the
introduction of a dislocation, and transforms 
into protected 1D gapless helical modes.

In the absence or presence of trivial ($b$: even) dislocations,
the finite-size gap evolves continuously 
into the bulk gap
as $s\rightarrow 0$.
When $b$ is odd, the same evolution 
gives {\it robustness} to the gapless modes.
When 
the circumference $s$ of the puncture, around which the crystal 
is dislocated, is finite,
the gapless modes are separated from the (gapped) continuum
only by an energy of order $A/s$.
As the size of the puncture is reduced,
only the gapless pairs 
stay intact, and its unique property that
it is topologically protected manifests,
making it 
distinguishable from the rest of the spectrum.
Projected Dirac cones without 
a pair of protected 1D gapless helical modes
become indistinguishable from the gapped bulk spectrum.

Fig. \ref{screw_c} (Fig. \ref{edge_c})
depicts such an evolution
in the presence of screw (edge) dislocations.
In the two figures, one can observe,
upon reducing the size of the cuts
(from lower to upper panels)
from $N_c =16$ either to $N_c =0$ (screw case) or to
$N_c =1$ (edge case),
that the gapless helical pair isolates.
Note that in the case of edge dislocations
one cannot reduce the cut width smaller than $b$.
Note also that in these plots separation between the two cuts
is relatively small ($N_y /2 =4$) in order to take
the width of the cut sufficiently large ($N_c =16$).
For this reason, there appears a finite interference
between the ideally gapless counter-propagating 
modes, each localized in the vicinity of 
two dislocation lines.
Of course,
when the size of the cut is finite, 
the wave function of the gapless mode
is extended almost uniformly around the cut.
The wave function shows a sharp peak in its amplitude 
in the vicinity of a dislocation line in the limit
$N_c \rightarrow 0$ (not shown).

\begin{figure}
\begin{center}
\includegraphics{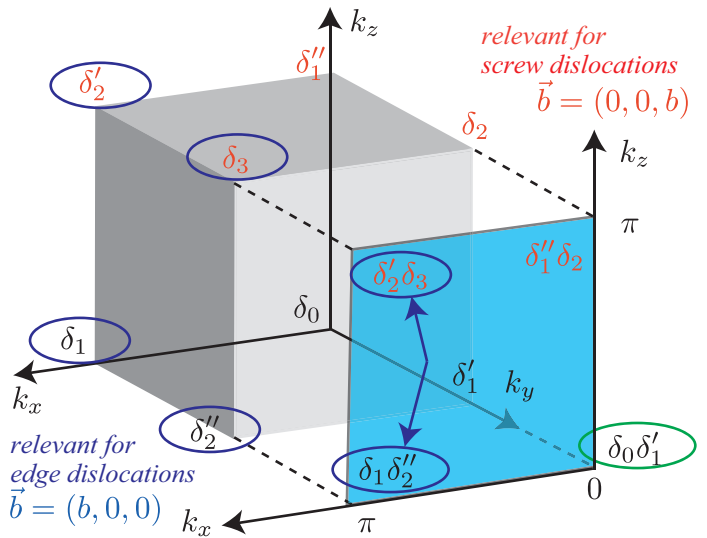}
\end{center}
\caption{Parity eigenvalues determining both the indices 
$(\nu_0; \nu_1 \nu_2 \nu_3)$ and
the position of surface Dirac cones on the projected 2D-plane;
here chosen as the $(k_z, k_x)$-plane.
Case 1: screw dislocation [$\vec{b} = (0, 0, b)$] ---
$\delta_{1}'' \delta_2$ and $\delta_{2}' \delta_3$
occurring at
$(k_z, k_x) = (\pi, 0)$ and $(k_z, k_x) = (\pi, \pi)$
are relevant.
Case 2: edge dislocation [$\vec{b} = (b, 0, 0)$] ---
$\delta_{1} \delta_{2}''$ and $\delta_{2}' \delta_3$
occurring at
$(k_z, k_x) = (0, \pi)$ and $(k_z, k_x) = (\pi, \pi)$
are relevant.}
\label{3D_2D}
\end{figure}

\begin{figure}
\begin{center}
\includegraphics{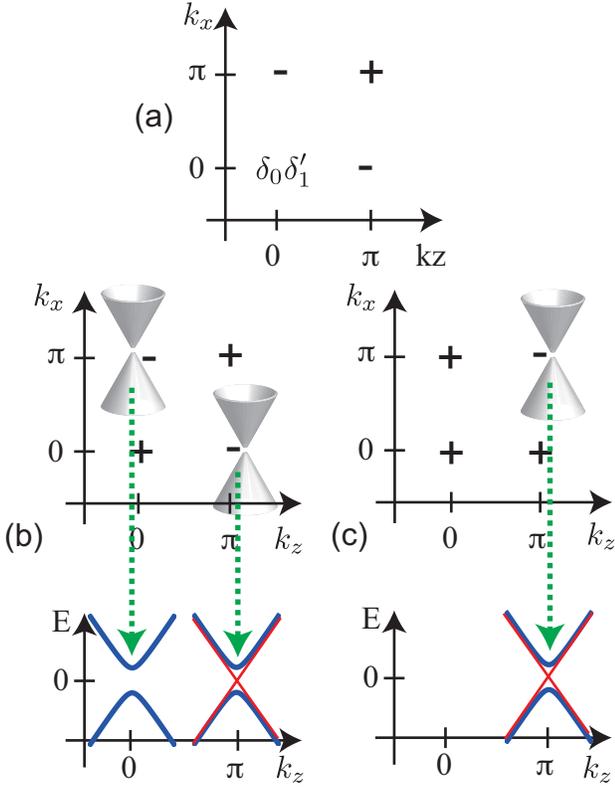}
\end{center}
\caption{(a) Necessary arrangement of
the 2D indices introduced in Fig. \ref{3D_2D}
for the appearance of protected 1D helical modes along 
a screw dislocation in the $z$-direction.
Only relative signs are relevant.
Column (b): WTI example 
satisfying the condition
in (a); $4<\Delta/B <8$.
Column (c): STI example 
satisfying the condition
in (a); $8<\Delta/B <12$. 
}
\label{2D_1D}
\end{figure}

\section{Relation between weak indices and
protected 1D helical modes}

What is the relation between the weak indices
and condition for the appearance of protected 1D helical modes?
A deep connection between these two a priori unrelated quantities
becomes manifest
when expressing both the weak indices and the latter condition
in terms of the parity eigenvalues at the 
eight bulk TRIM's, 
since our system has inversion symmetry.
\cite{Fu_Kane}
The expressions for weak indices  
in terms of $\delta_0$, $\delta_1$, $\delta_2$ and $\delta_3$
were given in Eq. (\ref{nu_3}).
In order to identify 
the condition for the appearance of protected 1D helical modes
in terms of $\delta_0$, $\delta_1$, $\delta_2$, $\delta_3$,
and $\delta$'s at 
symmetric but inequivalent points,
we project the 3D reciprocal space
in which the eight 3D TRIM are defined
in two steps;
first onto the 2D reciprocal surface on which surface Dirac cones
appear, 
then further onto 1D $k_z$-axis on which
protected 1D helical modes appear.
Fig. \ref{3D_2D} shows how the eight parity eigenvalues 
(among which only four are independent)
determine the weak indices, say, $\nu_3$,
upon projected onto the $(k_z, k_x)$-plane.
Products of two indices at four 2D TRIM
determine the position where surface Dirac cones appear.
\cite{Fu_Kane}

Fig. \ref{2D_1D} shows, on the other hand,
that the appearance or disappearance of protected 1D helical modes along 
a screw dislocation in the $z$-direction is
related to a relative sign of indices,
$\delta_1'' \delta_2$ and $\delta_2' \delta_3$
occurring at
$(k_z, k_x) = (\pi, 0)$ and $(k_z, k_x) = (\pi, \pi)$.
When these indices have opposite signs,
\begin{equation}
(\delta_1'' \delta_2) \times (\delta_2' \delta_3) = -1,
\label{surf_screw}
\end{equation}
there appears an odd number of, i.e., a single
2D surface Dirac cone that is projected to
$k_z = \pi$.
This projected Dirac cone acquires a finite-size mass gap
that is 
susceptible to the change of the boundary condition
(cf. Eq. (\ref{bc_2}))
caused by the twisting associated with, e.g., a screw dislocation.
The projected Dirac valley 
features a protected 1D helical modes when $b$ is an odd integer.
Notice, on the other hand, 
that the same combination of parity eigenvalues
as Eq. (\ref{surf_screw}) has appeared in Eq. (\ref{nu_3})
(see also Fig. \ref{3D_2D}).
Thus, "$\nu_3 =1$ and $b$ is odd" --- (A), is both 
a necessary and sufficient condition
for the appearance of protected 1D gapless helical modes.

The situation is different for an edge dislocation,
where the dislocation line is taken to be parallel to the $z$-axis 
but with a Burgers vector $\bm b = (\pm b, 0, 0)$.
In this case, 
the appearance of protected 1D modes is
related to a relative sign of the indices 
$\delta_1 \delta_2''$ and $\delta_2' \delta_3$
occurring at
$(k_z, k_x) = (0, \pi)$ and $(k_z, k_x) = (\pi, \pi)$;
see Fig. \ref{3D_2D}.
When these indices have opposite signs, i.e.,
\begin{equation}
(\delta_1 \delta_2'') \times (\delta_2' \delta_3) = -1,
\label{surf_edge}
\end{equation}
an odd number of surface Dirac cones are susceptible 
to the change of boundary condition (\ref{bc_3})
associated with 
the insertion or subtraction of crystal layers 
between the two cuts.
The same combination of $\delta's$ as Eq. (\ref{surf_edge})
has appeared, in contrast to the previous case,
in Eq. (\ref{nu_1}) (see also Fig. \ref{3D_2D}).
Thus protected 1D gapless modes appear in the present case
iff "$\nu_1 =1$ and $b$ is odd" --- (B).

The above statements (A) and (B) concerning the appearance of
protected 1D gapless modes
are consistent with the expression,
\begin{equation}
\vec{M}\cdot \vec{b} = \pi \ {\rm mod}\ 2\pi,
\label{M.b}
\end{equation}
which has appeared in Ref. \cite{Ran}.
This formula is a straight-forward generalization of the criteria
(A) and (B) to the case of the absence of inversion symmetry.
In Eq. (\ref{M.b})
the vector $\vec{M}$ is defined as,
\begin{equation}
\vec{M}={1\over 2}(\nu_1 \vec{G}_1+ \nu_2 \vec{G}_2 + \nu_3 \vec{G}_3),
\end{equation}
in terms of reciprocal lattice vectors,
$\vec{G}_1$, $\vec{G}_2$ and $\vec{G}_3$.
Note that the same formula can be derived by
considering winding properties of a Bloch electron
in an extended parameter space incorporating
the dislocation lines.
\cite{Teo_Kane}

\begin{figure}
\begin{center}
\includegraphics{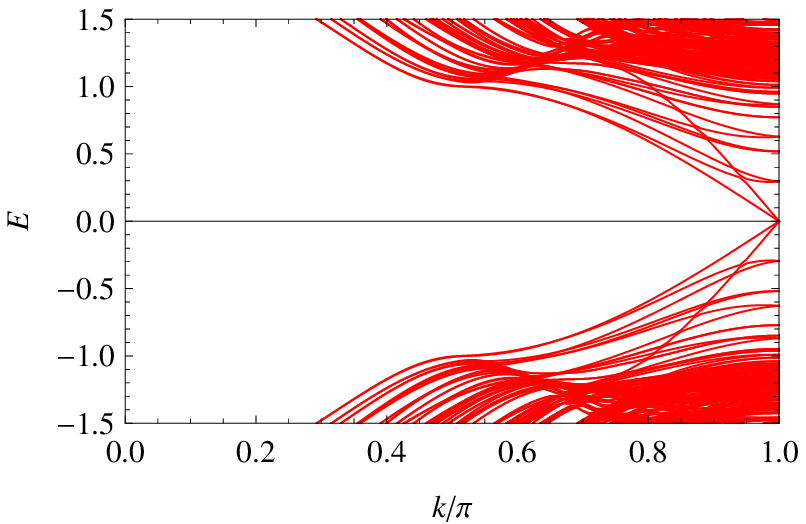}
\includegraphics{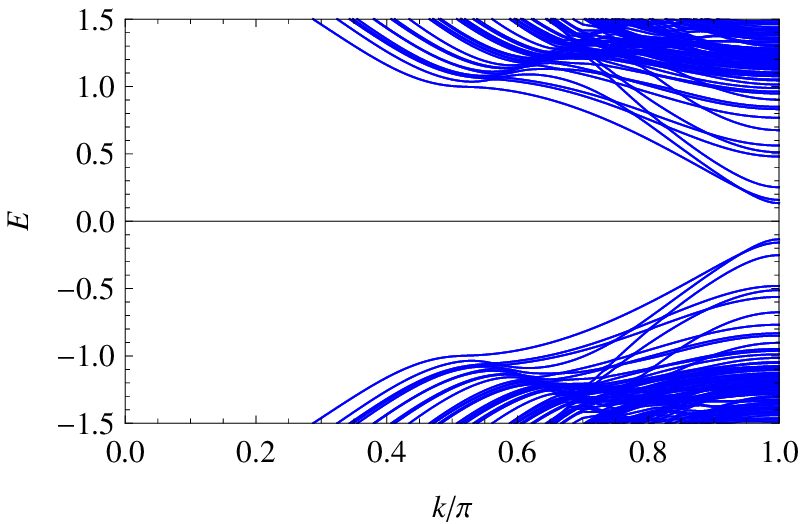}
\end{center}
\caption{Plots similar to Fig. \ref{spec_screw} and Fig. \ref{spec_edge}
here in the STI phase ($\Delta/B =10$).
The upper (lower) panel corresponds to the case of screw (edge)
dislocations, $b$ is chosen as, $b=3$ and $b=2$, respectively, 
in the two cases.
}
\label{spec_STI}
\end{figure}

In Fig. \ref{2D_1D} (a)
the lower-left index $\delta_0 \delta_1$ is irrelevant,
since Dirac cones projected onto $k_z =0$
is insensitive to the change of boundary condition
(cf. Eq. (\ref{bc_1})).
This means that 
the dislocation probes only weak indices.
Protected 1D gapless helical modes 
similarly appear
both in the WTI and STI phases with the same weak indices
(see Fig. \ref{2D_1D} columns (b-c)).

Fig. \ref{spec_STI} shows STI example on the (dis)appearance
of protected gapless modes, both in the screw and edge
dislocation cases.
Recall that in the WTI phase
protected gapless modes along an edge dislocation
appear at $k_z =0$,
whereas here the same protected modes
appear at $k_z =\pi$,
even though the two phases are characterized by the same weak indices;
$(\nu_0; \nu_1 \nu_2 \nu_3)=(0;111)$ [WTI: $4<\Delta/B<8$] and
$(\nu_0; \nu_1 \nu_2 \nu_3)=(1;111)$ [STI: $8<\Delta/B<12$].

\section{Conclusions}
In this paper, we have addressed the question: how ^^ ^^ weak" is 
a WTI?
The existence of protected gapless helical states 
parasitic to a dislocation line of a WTI seems per se 
contradictory to the fragility of the even numbers of surface Dirac cones 
of a WTI.
Using a simple model 
for a topological insulator implemented on a square lattice, 
we have systematically studied the nature of electronic states 
in the presence of dislocation lines.
In order to  
resolve the apparent contradiction 
between the stability of 1D gapless helical modes 
and the nonrobustness of 2D surface Dirac cones,
we have invented and studied a 
modified variant of the defect-free model in which a dislocation
is extended along a cylinder of finite circumference.
The unexpected stability of the 1D helical states was identified as
an interplay of the finite-size energy gap specific to
surface states of a 3D topological insulator and
twisting of the boundary condition 
due to topologically nontrivial geometry.
This scenario is closely related to the mechanism of 
recently observed anomalous Aharonov-Bohm oscillations
in a STI of ribbon geometry.

\acknowledgments
KI acknowledges Takahiro Fukui for stimulating discussions.
The authors are supported by KAKENHI; 
KI by Grant-in-Aid for Young Scientists (B) 19740189, 
KI and AT under the project on Innovative Areas,
``Topological Quantum Phenomena in Condensated Matter with Broken Symmetries''.


\bibliography{k1_r2_v5}


\end{document}